\documentclass[prd,aps]{revtex4}

\begin{document}

\title{Hyperbolicity of the BSSN system of Einstein evolution equations}

\author{Olivier Sarbach}
\author{Gioel Calabrese}
\author{Jorge Pullin}
\author{Manuel Tiglio}

\affiliation{Department of Physics and Astronomy, Louisiana State
University, 202 Nicholson Hall, Baton Rouge, Louisiana 70803-4001}

\begin{abstract}

We discuss an equivalence between the
Baumgarte--Shapiro--Shibata--Nakamura (BSSN) formulation of the
Einstein evolution equations, a subfamily of the Kidder--Scheel--Teukolsky
formulation, and other strongly
or symmetric hyperbolic first order systems with fixed shift and
densitized lapse. This allows us to
show under which conditions the BSSN system is, in a sense to be discussed,
hyperbolic. This
desirable property may account in part for the empirically observed
better behavior of the BSSN formulation in numerical evolutions
involving black holes.

\end{abstract}

\maketitle

\section{Introduction}

When numerically integrating the Einstein equations, it is common to
split them into constraint and evolution equations. Most
attempts usually solve
the ``free evolution'' problem, solving the constraints only at the
level of initial data. As in all gauge theories, the evolution
equations of general relativity are not uniquely defined. One can
redefine them by adding constraints, since the latter vanish on
solutions of the whole system of equations. It has been increasingly
recognized in the last few years (see \cite{Reula} for a list of
references), that it is desirable to take advantage of this freedom to
make the evolution equations strongly- (and preferably
symmetric-) hyperbolic. There is a body of mathematical literature
\cite{kreiss2} that one can draw from to prove stability (in the sense
of Lax's theorem) for discretization schemes based on well posed
formulations.  
In hyperbolic formulations one also has
better knowledge of the modes of the equations, something that is of
interest at the time of specifying boundary conditions that are
consistent with the constraints and the evolution equations \cite{boundaries1,boundaries2}.

The most traditional of the evolution schemes, the Arnowitt--Deser--Misner
(ADM) scheme, when written in first order form, is only weakly
hyperbolic (WH) \cite{kst}.
There has always been a feeling that this accounted in part for the 
difficulties encountered trying to evolve black holes and other spacetimes with this
evolution scheme. More recently \cite{convergence}, it was rigorously shown that popular 
discretizations of the ADM equations in particular, and WH
formulations in general, simply do not lead to convergent
schemes.

It was first recognized by Shibata and Nakamura \cite{SN} and further 
elaborated by Baumgarte and Shapiro \cite{BS} that if one performed
a conformal-traceless (CT) decomposition of the ADM variables, the
resulting scheme (commonly known as BSSN), at least in certain
situations, appeared to have less numerical difficulties than the ADM
system.   Also, anecdotal evidence suggested that slightly
different implementations of the BSSN scheme seemed to perform
quite differently in numerical simulations.

In this paper we would like to point out that the BSSN scheme, 
under certain conditions, is a reduction (from first to second order
in space)
of some strongly or symmetric hyperbolic systems. This may account in
part for its better performance in numerical simulations. The
conditions under which these reductions can be established 
depends rather subtly on details of the implementation
of the BSSN scheme, this may account for why some implementations 
appear better behaved than others. 

We will show that the BSSN scheme and its generalization by Alcubierre
et. al. \cite{alc1} can be mapped to the
Kidder--Scheel--Teukolsky (KST) \cite{kst}
system of evolution equations in the linearized regime. The latter is
a multi-parameter
family of evolution schemes, and we will discuss for which values of the
parameters the correspondence holds. The BSSN system is a first order in time 
and second order
in space formulation, whereas the KST system is first order in time and space. 
Therefore we will need to perform an enlargement of the BSSN system
to establish the correspondence. We will discuss in detail the 
precise nature of the correspondence below. We will show that a
similar enlargement exists at the nonlinear level,
which does not yield the KST system but which does yield a strongly or
symmetric hyperbolic one. The latter system is closely related to one
introduced by Friedrich and Rendall \cite{fri}.

We will start by presenting, in Section \ref{kst-bssn}, the KST 
system and summarizing their main properties.  Next we  present a 
CT decomposition of this system that shares its level of
hyperbolicity. We then 
discuss in detail the correspondence in linearized gravity between this
CT decomposition of the
KST system and the BSSN equations. In Section \ref{bssn-nonlin} we show that by
introducing extra variables to make the BSSN system first order in
time and space one can get strongly and symmetric hyperbolic systems provided that
the lapse is appropriately densitized.
We end with a discussion of the relevance of these results for numerical
simulations.

%%%%%%%%%%%%%%%%%%%%%%%%%%%%%%%%%%%%%%%%%%%%%%%%%%%%%%%%%%%%%%%%%%%
\section{The BSSN-KST correspondence}
\label{kst-bssn}
%%%%%%%%%%%%%%%%%%%%%%%%%%%%%%%%%%%%%%%%%%%%%%%%%%%%%%%%%%%%%%%%%%%

In this section, we show that for the case of linearization around
flat space-time in Cartesian coordinates the
BSSN equations are equivalent (in a sense to be defined below) to 
the linearized KST equations with an appropriate choice of
parameters. As a byproduct of our construction, we present a CT
decomposition of the KST system that shares its same level of
hyperbolicity, even at the nonlinear level.  

%%%%%%%%%%%%%%%%%%%%%%%%%%%%%%%%%%
\subsection{The KST family of equations}
%%%%%%%%%%%%%%%%%%%%%%%%%%%%%%%%%%

Starting from the ADM equations, KST 
derive a family of strongly  and symmetric hyperbolic 
first-order evolution equations
for the three-metric ($g_{ij}$), the extrinsic curvature ($K_{ij}$), and the
spatial derivatives of the three-metric ($d_{kij} \equiv \partial_k
g_{ij}$), which generalizes previous well posed formulations \cite{fr,ec}.
A priori prescribing  the densitized
lapse $\exp{(Q)}$ as a function of spacetime, the lapse $N$ is given by
\begin{equation}
N = g^\sigma e^Q, \label{densitizing_lapse}
\end{equation}
where $g$ is the
determinant of the three-metric and $\sigma$ a parameter. 
The shift vector, $\beta^i$, is also assumed to be prescribed as a
function of spacetime. Here we
use the same notation as in KST except that
 we define the derivative operator along the normal to the
hypersurfaces as
\begin{displaymath}
\partial_0 = \frac{1}{N} (\partial_t - \pounds_\beta).
\end{displaymath}

The vacuum evolution equations have the form
\begin{eqnarray}
\partial_0 g_{ij} &=& - 2K_{ij}\, , \label{Eq:KST1}\\
\partial_0 K_{ij} &=& - \frac{1}{2} \partial^k d_{kij}
  + \frac{1}{2}\partial^k d_{(ij)k} + \frac{1}{2} g^{ab} \partial_{(i} d_{|ab|j)}
  - \left(\frac{1}{2} + \sigma \right) g^{ab}\partial_{(i} d_{j)ab}
\nonumber\\
 &+& \zeta\, g^{ab} C_{a(ij)b}
  + \gamma\, g_{ij} C - e^{-Q} \partial_i\partial_j (e^Q) + {\cal R}_{ij}\, ,
\label{Eq:KST2}\\
\partial_0 d_{kij} &=& - 2\partial_k K_{ij}
  + \eta\, g_{k(i} C_{j)} + \chi\, g_{ij} C_k + {\cal R}_{kij}\, ,
\label{Eq:KST3}
\end{eqnarray}
where the constraint variables
\begin{eqnarray}
& C = \frac{1}{2} g^{ab}\partial^k (d_{abk} - d_{kab}) + {\cal R}, & \hbox{(Hamiltonian constraint)}
\nonumber\\
& C_j = \partial^a K_{aj} - g^{ab}\partial_j K_{ab} + {\cal R}_j, & \hbox{(momentum constraint)}
\nonumber\\
& C_{kij} = d_{kij} - \partial_k g_{ij}\, , & \hbox{(definition of $d_{kij}$)}
\nonumber\\
& C_{lkij} = \partial_{[l} d_{k]ij}\, . & \hbox{(''closedness'' of $d_{kij}$)}
\nonumber
\end{eqnarray}
have been added to the right-hand side of the evolution equations
with some free parameters $\zeta , \gamma , \eta, \chi$.
Here, $\partial^k \equiv g^{kl} \partial_l\,$ and
$b_j$, $d_k$ are defined as the traces of $d_{kij}\,$:
\begin{displaymath}
b_j = g^{ki} d_{kij}\, , \;\;\;
d_k = g^{ij} d_{kij}\, .
\end{displaymath}
The ``remainder'' terms ${\cal R}$, ${\cal R}_i$, ${\cal R}_{ij}$ and ${\cal R}_{kij}$
are homogeneous polynomials of degree $2$ in $d_{kij}$, $\partial_i Q$
and $K_{ij}$, where the coefficients may depend on $g_{ij}$.
Finally, the Lie derivative of the symbols $d_{kij}$ is
\begin{displaymath}
\pounds_\beta d_{kij} = \beta^l\partial_l d_{kij} + d_{lij}\partial_k\beta^l
 + 2d_{kl(i}\partial_{j)}\beta^l + 2 g_{l(i} \partial_{j)}\partial_k\beta^l.
\end{displaymath}

The evolution system has characteristic speeds
$\{ 0, \pm 1, \pm\sqrt{\lambda_1}, \pm\sqrt{\lambda_2}, \pm\sqrt{\lambda_3} \}$, 
where
\begin{eqnarray}
\lambda_1 &=& 2\sigma,
\nonumber\\
\lambda_2 &=& 1 + \chi - \frac{1}{2}(1 + \zeta)\eta + \gamma(2 - \eta + 2\chi),
\nonumber\\
\lambda_3 &=& \frac{1}{2}\,\chi + \frac{3}{8}(1 - \zeta)\eta
     -\frac{1}{4}(1 + 2\sigma)(\eta + 3\chi). \nonumber
\end{eqnarray}
It should be noted that these are the characteristic speeds with
respect to the $\partial _0$ operator. This means that $\lambda =0,1$
correspond to propagation along the normal to the hypersurfaces or the
light cone, respectively (physical speeds). The characteristic speeds with respect to
the $\partial_t$ operator are obtained from these after the
transformation
\begin{displaymath}
\mu \mapsto N\mu + \beta^i n_i\; ,
\end{displaymath}
where $n^i$ is the direction of the corresponding characteristic mode.

The conditions under which the system is completely ill posed (CIP),
WH or strongly hyperbolic (SH)  were found by KST.
The system is CIP if any of the above speeds is complex,
while it is WH if $\lambda_j \geq 0$ for $j=1,2,3$ but one of the
conditions (\ref{eq:SH1}, \ref{eq:SH2}) below is violated.
For example, if the parameters ($\zeta$, $\gamma$, $\eta$,
$\chi$) are zero the dynamics is equivalent to the ADM equations 
 written in first order form with fixed densitized lapse and fixed
shift (which are WH). If $\sigma =0$ as well, then the system is
equivalent to the ADM equations with fixed lapse and shift (which are
also WH). 

Finally, the system is SH provided that
\begin{eqnarray}
&& \lambda_j > 0, \;\;\; \hbox{for $j=1,2,3$}, \label{eq:SH1}\\
&& \lambda_3 = \frac{1}{4}(3\lambda_1 + 1) \;\;\;
  \hbox{if $\lambda_1 = \lambda_2$}. \label{eq:SH2}
\end{eqnarray}
Here by SH it is meant that the principal part has a complete set of
 eigenvectors with real
eigenvalues. Provided there is a smooth and uniformly bounded
symmetrizer, this ensures well posedness \cite{kreiss2}. 

In KST seven extra parameters are introduced and used to make
changes of variables in $K_{ij}$ and $d_{kij}$. When performing this
change of variables the constraint $C_{kij}=0$ is also used in order to
trade spatial derivatives of the three-metric for $d_{kij}$.
Thus, the equations with the new variables have different
solutions off the constraint surface.
However, one can see that at the linear level
this change of variables does not involve the addition of
constraints.

%%%%%%%%%%%%%%%%%%%%%%%%%%%%%%%%%%%%%%%%%%%%%%%%%%%%%%
\subsection{Conformal trace-less decomposition of the KST system}
%%%%%%%%%%%%%%%%%%%%%%%%%%%%%%%%%%%%%%%%%%%%%%%%%%%%%%

We start by performing a CT of the KST
variables:
\begin{eqnarray}
g_{ij} &=& e^{4\phi} \tilde{g}_{ij}, \nonumber\\
K_{ij} &=& e^{4\phi} \left( A_{ij} + \frac{1}{3} \tilde{g}_{ij} K \right),
\nonumber\\
d_{kij} &=& e^{4\phi} \left( 2e_{kij} + \frac{3}{5} \tilde{g}_{k(i} \Gamma_{j)}
 - \frac{1}{5} \tilde{g}_{ij} \Gamma_k + \frac{1}{3}\tilde{g}_{ij} d_k \right).
\nonumber
\end{eqnarray}
Here $\tilde{g}_{ij}$ has unit determinant, $A_{ij}$ is
trace-less and $e_{kij}$ is trace-less over all pairs of indices, and 
$\Gamma_j$ is defined by $\Gamma_j = b_j - d_j/3$. In terms of the
Christoffel symbols $\tilde{\Gamma}^k_{\; ij}$ belonging to the conformal
metric $\tilde{g}_{ij}$, $\Gamma_j$ can also be expressed as 
($\tilde{\partial}^i \equiv \tilde{g}^{ik} \partial_k\, $)
\begin{displaymath}
\Gamma_j = \tilde{g}^{ki} \tilde{\Gamma}_{jki} = \tilde{\partial}^i \tilde{g}_{ij}\, ,
\end{displaymath}
which is the additional variable used in the BSSN system
(actually, in the original version by Shibata and Nakamura, the variable
$F_j = \delta^{ki}\partial_k \tilde{g}_{ij}$ is used instead of $\Gamma_j$,
in the version by Baumgarte and Shapiro, $\Gamma^k = \tilde{g}^{kj} \Gamma_j$
is used.)

With respect to this decomposition, the KST system (\ref{Eq:KST1}-\ref{Eq:KST3})
becomes, up to lower order terms (``$l.o.$'')
quadratic in $d_{kij}$, $\partial_i Q$ and $K_{ij}$,
\begin{eqnarray}
\partial_0\phi &=& - \frac{1}{6}\, K, \label{Eq:KSTCT1}\\
\partial_0\tilde{g}_{ij} &=& - 2 A_{ij}\, ,\label{Eq:KSTCT2}\\
\partial_0 K &=& \left( 1 + \frac{3}{2}\gamma \right) \partial^k\Gamma_k
  - \left( \frac{2}{3} + \gamma + \sigma \right) \partial^k d_k - e^{-Q}\partial^k\partial_k (e^Q) 
+ l.o.\, , \label{Eq:KSTCT3}\\
\partial_0 A_{ij} &=& - \partial^k e_{kij} + (1 + \zeta)\partial^k e_{(ij)k}
  + e^{-4\phi} \left[  \frac{1}{20}(5 - 9\zeta) \partial_{(i} \Gamma_{j)} - \left( \frac{1}{6} + \sigma \right) \partial_{(i} d_{j)}
  - e^{-Q} \partial_i\partial_j (e^Q) \right]^{TF} + (l.o.)_{ij}\, ,\quad \label{Eq:KSTCT4}\\
\partial_0 d_k &=& (\eta + 3\chi)\tilde{\partial}^s A_{sk}
 - \frac{2}{3}\left( \eta + 3\chi + 3 \right) \partial_k K + (l.o.)_k\, , \label{Eq:KSTCT5}\\
\partial_0\Gamma_j &=& \left( \frac{5}{3}\eta - 2\right)\tilde{\partial}^s A_{sj} - \frac{10}{9}\eta\,\partial_j K + (l.o.)_j\, ,
\label{Eq:KSTCT6}\\
\partial_0 e_{kij} &=& = - \partial_k A_{ij}
 + \frac{3}{5} \tilde{g}_{k(i} \tilde{\partial}^s A_{j)s} - \frac{1}{5}\tilde{g}_{ij} \tilde{\partial}^s A_{sk} + (l.o.)_{kij}\, ,
\label{Eq:KSTCT7}
\end{eqnarray}
where the superscript $TF$ denotes the trace-free part.
Here, for simplicity, we have also assumed that the shift
vanishes. However, a 
generalization in the presence of shift can also be worked out. 

Notice that in the derivation of Eqs. (\ref{Eq:KSTCT1}-\ref{Eq:KSTCT7})
we have used the  constraints $C_{kij} = 0$ in order to re-express
first order spatial derivatives of $g_{ij}$ in terms of $d_{kij}$ in a
way that ensures that the principal part of the transformed equations
has the same spectrum as the original KST system.
Therefore, for a given choice of parameters both systems have the same level of
hyperbolicity. However, in general the dynamics of the evolution Eqs. 
(\ref{Eq:KSTCT1}-\ref{Eq:KSTCT7}) is not identical to the dynamics of the KST 
Eqs. (\ref{Eq:KST1}-\ref{Eq:KST3})
if we are off the constraint surface $C_{kij} = 0$.
A similar situation occurs in KST when the variables $K_{ij}$
and $d_{kij}$ are redefined through a seven-parameter family of change
of variables.

%%%%%%%%%%%%%%%%%%%%%%%%%%%%%%%%%%%%%%%%%%%%%%%%%%%%%%%%
\subsection{Correspondence}
%%%%%%%%%%%%%%%%%%%%%%%%%%%%%%%%%%%%%%%%%%%%%%%%%%%%%%%

We start by looking at the linearized equations 
around flat spacetime,
written in Cartesian coordinates. One can see that in this case all lower order
terms (``$l.o.$'') vanish, and the two systems (\ref{Eq:KST1}-\ref{Eq:KST3})
and (\ref{Eq:KSTCT1}-\ref{Eq:KSTCT7}) are dynamically equivalent
(i.e. have the same set of solutions, even off the constraint surface).
In the following, we will call the linearized version of equations
(\ref{Eq:KSTCT1}-\ref{Eq:KSTCT7}) system A. 

Since the BSSN system has second order spatial derivatives and the KST
system has only first order spatial derivatives, care needs to be 
exercised concerning the constraints that materialize the correspondence
of the spatial derivatives. Basically, we need to show that if we start
with initial data that is on the submanifold of constraints that 
relate the second spatial derivatives to first spatial derivatives
of extra variables (all solutions of the BSSN system automatically
lie on this submanifold), we automatically remain on it. Otherwise,
no correspondence will be possible.

Consider the constraints
\begin{displaymath}
\tilde{C}_{kij} = e_{kij} - \frac{1}{2} \partial_k \tilde{g}_{ij} 
  + \frac{3}{10} \delta_{k(i} \partial^s \tilde{g}_{j)s} 
  - \frac{1}{10} \delta_{ij} \partial^s \tilde{g}_{ks} \, .
\end{displaymath}
Using the linearized version of evolution equations
(\ref{Eq:KSTCT2}) and (\ref{Eq:KSTCT7}),
one can see that these constraints are constant in time,
\begin{displaymath}
\partial_0 \tilde{C}_{kij} = 0,
\end{displaymath}
independently of whether or not the remaining constraints are
satisfied. The constraints $\tilde{C}_k = d_k - 12\partial_k\phi$ are constant in
time as
well, provided that $\eta + 3\chi = 0$.
Therefore, if we give initial data with $\tilde{C}_{kij} = 0$ and
$\tilde{C}_k = 0$, the equations
\begin{eqnarray}
e_{kij} &=& \frac{1}{2} \partial_k \tilde{g}_{ij} 
  - \frac{3}{10} \delta_{k(i} \partial^s \tilde{g}_{j)s} 
  + \frac{1}{10} \delta_{ij} \partial^s \tilde{g}_{ks}, \label{Eq:ekijDef}\\
d_k &=& 12\partial_k\phi, \label{Eq:dkDef}
\end{eqnarray}
hold at all times, provided that $\eta + 3\chi = 0$.
In this case, we can replace $e_{kij}$ and $d_k$ by the right-hand side
of equations (\ref{Eq:ekijDef}) and (\ref{Eq:dkDef}) everywhere they
appear in the linearized version of equations
(\ref{Eq:KSTCT1}-\ref{Eq:KSTCT4},\ref{Eq:KSTCT6})
and forget about the evolution equations for the $15$ variables
$e_{kij}$, $d_k$. The evolution equations for $K$ and $A_{ij}$ become
\begin{eqnarray}
\partial_0 K &=& \left( 1 + \frac{3}{2}\gamma \right) 
\partial^k\Gamma_k  - 8\left( 1 + \frac{3}{2}\,\gamma + \frac{3}{2}\,\sigma \right) 
\partial^k\partial_k\phi - e^{-Q} \partial^k\partial_k (e^Q),
\label{Eq:LinBSSN1}\\
\partial_0 A_{ij} &=& - \frac{1}{2} \partial^k\partial_k \tilde{g}_{ij} + \left[ \partial_{(i} \Gamma_{j)} 
 + \frac{1}{20}(15 + 9\zeta)\left( \partial_{(i}\partial^s \tilde{g}_{j)s} - \partial_{(i} \Gamma_{j)} \right) 
 - 2(1 + 6\sigma) \partial_i\partial_j\phi - e^{-Q} \partial_i\partial_j (e^Q) \right]^{TF}.
\label{Eq:LinBSSN2}
\end{eqnarray}
In the following, we will call the system which constitutes the linearized
version of equations (\ref{Eq:KSTCT1},\ref{Eq:KSTCT2},\ref{Eq:KSTCT6}) and
the equations (\ref{Eq:LinBSSN1},\ref{Eq:LinBSSN2}) system B.
If initial data is chosen for system B and one takes the same initial data
for system A with $d_k$ and $e_{kij}$ given by equations (\ref{Eq:ekijDef},\ref{Eq:dkDef}),
the two systems will yield the same result upon evolution (if there are no boundaries
or if we have time-like boundaries to which $\partial_0$ is tangential.)

If we compare system B with the generalization of the BSSN
system  by Alcubierre {\it et al}. \cite{alc1}  in the linearized regime, we see
that the two agree if the KST parameters are
\begin{equation}
\zeta = -\frac{5}{3}\, , \;\;\;
\gamma = -\frac{2}{3}\, ,\;\;\;
\eta = \frac{6}{5} m\, ,\;\;\;
\chi = -\frac{2}{5} m\, .
\label{Eq:BSSNparam}
\end{equation}
Notice that since \cite{alc1} reduces to BSSN \cite{SN,BS} when $m=1$,
system B agrees with the original BSSN as well for these values of the
parameters. Notice also that for the equivalence to be true one has to
use in BSSN the same gauge conditions used in KST
(Eq. (\ref{densitizing_lapse}) with arbitrary $\sigma $).

For example, the choice $\gamma = -\frac{2}{3}$ eliminates the
Ricci scalar appearing in the evolution equation for $K$, and
choosing $\eta = \frac{6}{5}$ eliminates the term $\partial^s A_{sk}$
in the evolution equation for the variable $\Gamma_j$,
as in the traditional BSSN system (while in \cite{alc1} the parameter $m$
is allowed to vary).
Therefore, the linearized BSSN equations are dynamically equivalent
to the linearized KST system with the choice of parameters given by
(\ref{Eq:BSSNparam}), provided that the constraints (\ref{Eq:ekijDef},\ref{Eq:dkDef})
are satisfied. In particular, any solution to the linearized BSSN
equations (even one that does not satisfy the Hamiltonian or momentum
constraint, or the constraint that corresponds to the definition of
$\Gamma_j$) gives a solution to the linearized KST system with
(\ref{Eq:BSSNparam}) if we define $d_k$ and $e_{kij}$ according
to (\ref{Eq:ekijDef},\ref{Eq:dkDef}).

Now an interesting question is whether or not the KST subfamily
that corresponds to the linearized BSSN system is SH.
Recalling Section \ref{kst-bssn}, it turns out that the 
characteristic speeds are plus or minus the square root of
\begin{displaymath}
\lambda_1 = 2\sigma , \;\;\;
\lambda_2 = \frac{1}{3}(4m - 1), \;\;\;
\lambda_3 = m, \;\;\;
\lambda_4 = 1.
\end{displaymath}
In particular, we see that densitizing the lapse with $\sigma >0 $ in 
(\ref{Eq:BSSNparam}) corresponds to a SH system 
as long as $m > 1/4$. In particular, if $\sigma = 1/2$ and $m=1$,
the system has only physical speeds ($\lambda = 0,1$). It is also
worthwhile pointing out that in this (linear, constant coefficient)
case well posedness can be shown for the SH families, since all is needed
is the boundedness of the symmetrizer, which can be shown to hold.

One might wonder whether the parameters (\ref{Eq:BSSNparam}) 
(with $\sigma = 1/2$) correspond to the ones found empirically by KST
to have better stability properties.
The answer is no, since the one-parameter subfamily (\ref{Eq:BSSNparam}) 
has no intersection with the two-parameter Einstein-Christoffel (EC) 
\cite{ec} generalization used in the numerical simulations
of KST.

%%%%%%%%%%%%%%%%%%%%%%%%%%%%%%%%%%%%%%%%%%%%%%%%%%%%%%%%%%%%%%
\section{Hyperbolic first order enlargements of the BSSN system}
\label{bssn-nonlin}
%%%%%%%%%%%%%%%%%%%%%%%%%%%%%%%%%%%%%%%%%%%%%%%%%%%%%%%%%%%%%%

We were not able to find a way in the non-linear case to get rid of the
extra variables $e_{kij}$ and $d_k$ when lower order terms are present
in the equations. We will therefore not prove a 
correspondence between the KST and BSSN systems but, instead, a correspondence
between the BSSN system and certain SH or symmetric hyperbolic formulations
closely related to one by Friedrich and Rendall \cite{fri}. 
Starting with the (non-linear) BSSN system with vanishing shift (see
below for the case of non-vanishing shift)
we are able to show the following result: by introducing $18$ extra variables, the BSSN
system can be written as a system which is first order in time and space
and whose principal part looks very similar to the one of the CT decomposition of 
the KST system. Drawing from our experience
with the latter, one can then show that the new
first order system has the same characteristic speeds as system B
with the choice of parameters (\ref{Eq:BSSNparam}). 
With a specific choice of parameters $\sigma$ and $m$ this system has
the same level of hyperbolicity as the KST one.

The non-linear BSSN equations with zero shift have the form
\begin{eqnarray}
\partial_0\phi &=& -\frac{1}{6} K, \label{Eq:BSSN1}\\
\partial_0 \tilde{g}_{ij} &=& -2 A_{ij}\, , \label{Eq:BSSN2}\\\
\partial_0 K &=& -12\sigma \partial^k\partial_k\phi - e^{-Q} \partial^k\partial_k (e^Q) + l.o.\, , \label{Eq:BSSN3}\\\
\partial_0 A_{ij} &=& e^{-4\phi} \left[ -\frac{1}{2} \tilde{\partial}^k\partial_k \tilde{g}_{ij} + \partial_{(i} \Gamma_{j)}
  - 2(1 + 6\sigma)\partial_i\partial_j\phi - e^{-Q} \partial_i\partial_j (e^Q) \right]^{TF} + (l.o.)_{ij}\, , \label{Eq:BSSN4}\\\
\partial_0 \Gamma_i &=& 2(m-1)\tilde{\partial}^s A_{si} - \frac{4m}{3} \partial_i K + (l.o.)_i\, , \label{Eq:BSSN5}\
\end{eqnarray}
where $\partial_0 = \partial_t/N$, and now ``$l.o.$'' denote terms that contain
only derivatives of $\phi$, $\tilde{g}_{ij}$, $Q$ up to first order,
and no derivatives of the other variables.
Here, the lapse $N$ and the densitized lapse $\exp{(Q)}$ are related
to each other as in equation (\ref{densitizing_lapse}).
In terms of the BSSN variables,
\begin{equation}
N = e^{12\sigma\phi + Q}.
\end{equation}
For generality, we have also added the momentum constraint to the
evolution equation for $\Gamma_i$ with an arbitrary factor $m$ as in
Alcubierre et. al. \cite{alc1}.

Our aim is to recast this system of equations into a first order
system by introducing extra
variables,
\begin{equation}
d_k = 12\partial_k\phi, \;\;\;
\tilde{d}_{kij} = \partial_k \tilde{g}_{ij}\, ,
\label{Eq:ExtraVars}
\end{equation}
 and using only the constraints that are associated to
the definition of these extra variables. Using the fact that
\begin{displaymath}
[\partial_0,\partial_k] = \frac{N_k}{N}\partial_0 = \left( 12\sigma\partial_k\phi + \partial_k Q \right)\partial_0
 = \left( \sigma d_k + \partial_k Q \right)\partial_0\, ,
\end{displaymath}
and Eqs. (\ref{Eq:BSSN1},\ref{Eq:BSSN2}),
we obtain the following evolution equations for $d_k$ and
$\tilde{d}_{kij}$:
\begin{eqnarray}
\partial_0 d_k &=& -2\left(\partial_k + \sigma d_k + \partial_k Q \right)K, 
\label{Eq:Evoldk}\\
\partial_0\tilde{d}_{kij} &=& -2\left(\partial_k + \sigma d_k + \partial_k Q \right)A_{ij}\, .
\label{Eq:Evoldkij}
\end{eqnarray}
The constraints associated to 
the definition of $d_k$ and $\tilde{d}_{kij}$,  $\tilde{C}_k$ and
$\tilde{C}_{kij}$, satisfy
\begin{eqnarray}
\partial_0 \tilde{C}_k &=& -2\sigma K \tilde{C}_k\, ,\\
\partial_0 \tilde{C}_{kij} &=& -2\sigma A_{ij} \tilde{C}_k , 
\end{eqnarray}
independently on whether or not the remaining constraints are satisfied.
Since the evolution equations for $\tilde{C}_k$ and $\tilde{C}_{kij}$ form
a closed system, the equations that define $d_k$ and $\tilde{d}_{kij}$
are satisfied automatically during evolution (provided that they are satisfied
initially and  appropriate conditions 
are given at time-like boundaries.)

Thus, we can enlarge system (\ref{Eq:BSSN1}-\ref{Eq:BSSN5}) by adding
the evolution equations (\ref{Eq:Evoldk},\ref{Eq:Evoldkij}) above
and by replacing $12\partial_k\phi$ by $d_k$ and $\partial_k \tilde{g}_{ij}$ by
$\tilde{d}_{kij}$, respectively. The resulting system is
\begin{eqnarray}
\partial_0\phi &=& - \frac{1}{6}\, K, \label{Eq:BSSNF1}\\
\partial_0 \tilde{g}_{ij} &=& - 2 A_{ij}\, ,\label{Eq:BSSNF2}\\
\partial_0 K &=& -\sigma \partial^k d_k - e^{-Q} \partial^k\partial_k (e^Q) + l.o.\, , \label{Eq:BSSNF3}\\
\partial_0 A_{ij} &=& -\frac{1}{2}\partial^k \tilde{d}_{kij} 
  + e^{-4\phi} \left[ \partial_{(i} \Gamma_{j)} - \left( \frac{1}{6} + \sigma \right) \partial_{(i} d_{j)}
  - e^{-Q} \partial_i\partial_j (e^Q) \right]^{TF} + (l.o.)_{ij}\, , \label{Eq:BSSNF4}\\
\partial_0\Gamma_j &=& 2(m-1)\tilde{\partial}^s A_{sj} - \frac{4m}{3}\partial_j K + (l.o.)_j\, ,
\label{Eq:BSSNF5}\\
\partial_0 d_k &=& -2\partial_k K + (l.o.)_k, \label{Eq:BSSNF6}\\
\partial_0 \tilde{d}_{kij} &=& -2\partial_k A_{ij} + (l.o.)_{kij}\, ,
\label{Eq:BSSNF7}
\end{eqnarray}
where after the replacements mentioned above, no derivatives of the
dynamical variables appear in ``$l.o.$''. 
Analyzing the level of hyperbolicity, it turns out
that the system (\ref{Eq:BSSNF1}-\ref{Eq:BSSNF7}) has the same
characteristic speeds as the KST system with the choice
(\ref{Eq:BSSNparam}) and the same level of hyperbolicity. In
particular, it is SH if $\sigma >0 $ and $m>1/4$, and
has only physical characteristic speeds if $\sigma =1/2$ and $m=1$.  

Furthermore, it is easy to see that if the relation
\begin{equation}
2\sigma = \frac{1}{3}(4m - 1)
\label{Eq:SymHypRel}
\end{equation}
holds and $m > 1$, the principal part of the system is symmetric  
with respect to the inner product associated with 
\begin{displaymath}
(u,u) \equiv \phi^2 + \tilde{g}^{ij} \tilde{g}_{ij} + \frac{3}{4m-1}\, K^2
  + A^{ij} A_{ij} + \frac{e^{-4\phi}}{2(m-1)}\left( \Gamma^k \Gamma_k
  - \frac{4m}{3} \Gamma^k d_k
  + \frac{4 m^2}{9} d^k d_k \right)
  + \frac{e^{-4\phi}}{4} \left( d^k d_k + \tilde{d}^{kij}
  \tilde{d}_{kij} \right)\, , \quad
\end{displaymath}
where $u = (\phi,\tilde{g}_{ij}, K, A_{ij}, d_k, \Gamma_j, \tilde{d}_{kij})^T$
and where indices are raised with the metric $\tilde{g}_{ij}$. A
point to be remarked is that this symmetric hyperbolic subfamily has
superluminal speeds, since we have to impose the condition $m>1$ and
Eq. (\ref{Eq:SymHypRel}) must hold. In particular, we cannot recover
the original BSSN
system ($m=1$) but, instead, its generalization by
Alcubierre et. al..

To summarize, we have shown that any version of BSSN which has the form
(\ref{Eq:BSSN1}-\ref{Eq:BSSN5}) can be rewritten as a first order system
in time and space by introducing the extra variables (\ref{Eq:ExtraVars}).
Provided that initially the constraints associated to the definition
of the new variables
are satisfied, this first order system is equivalent to the BSSN system
chosen. For a fixed densitized lapse, the system is SH
if $\sigma > 0$ and $m > 1/4$. If $m > 1$ and we choose to densitize
the lapse according to Eq. (\ref{Eq:SymHypRel}), 
the system is even symmetric hyperbolic.
At this point it is worthwhile pointing out that a similar enlargement
to a SH first order system does not work in the ADM case,
even after densitizing the lapse. In order to get a SH system,
one must add the  Hamiltonian or momentum constraints to the evolution equations.

If the shift is non-trivial, one can still obtain the above
enlargement. But in this case, divergence terms of the conformal
metric $\tilde{g}_{ij}$ have to be substituted by $\Gamma_i$'s in such
a way that the equations have the form (\ref{Eq:BSSN1}-\ref{Eq:BSSN5})
with $\partial_0 = (\partial_t - \beta^k\partial_k)/N$ and where the
lower order terms (``$l.o.$'') may depend on $\beta^i$. In this case, 
the enlargement is SH or symmetric hyperbolic under the same
conditions as is the zero shift case provided the shift is a 
prescribed function.

Frittelli and Reula (FR) have also presented well posed CT 
formulations of Einstein's
equations \cite{fr2} that have attractive properties (for example, only
physical characteristic speeds), but they do not seem
to reduce to the BSSN system in
the sense described above due to the way the constraints are added in
their construction.

%%%%%%%%%%%%%%%%%%%%%%%%%%%%%%%%%%%%
\section{Summary}
%%%%%%%%%%%%%%%%%%%%%%%%%%%%%%%%%%
We have shown that at the linearized level the BSSN system with fixed
densitized lapse and fixed shift is
equivalent to a SH subfamily of the KST formulation, while at the
nonlinear level it is equivalent to certain SH or symmetric hyperbolic
system. This equivalence is analogous 
 to the one between the wave equation written as first order
in time {\em and} space, or first order in time but second order in
space. 

The hyperbolicity of BSSN (and any other system) depends delicately
on how the equations are written. Small substitutions of 
some equations in others are enough to spoil hyperbolicity.
This may account for several anecdotal reports that rewriting
part of the BSSN equations appears to improve or worsen 
the numerical behavior. Having established that this system
is  hyperbolic opens the possibility of rigorously
showing that the resulting numerical schemes constructed 
upon discretization are stable in the sense of Lax's theorem,
and therefore the resulting codes are convergent. This property
is not available at present for WH formulations,
including the ADM equations or even small modifications of the BSSN
system considered here. Another advantage is that one
could also derive consistent boundary conditions for the BSSN
equations or show well posedness by looking at the
first order hyperbolic enlargement and later reducing the result back
to second order in space. 
This therefore lays mathematical grounds on which to understand and
make use of the 
advantages in numerical implementations of the BSSN equations.

%%%%%%%%%%%%%%%%%%%%%%%%%%%%%%%%%%%%%%%%%%%%%%%%%%%%%%%%%%%%%%
\section{Acknowledgments}
%%%%%%%%%%%%%%%%%%%%%%%%%%%%%%%%%%%%%%%%%%%%%%%%%%%%%%%%%%%%%%
This work was supported in part by  grant NSF-PHY9800973, by the
Horace C. Hearne Jr. Institute of Theoretical Physics, the Swiss National Science
Foundation, and Fundaci\'on Antorchas.

%%%%%%%%%%%%%%%%%%%%%%%%%%%%%%%%%%%%%%%%%%%%%%%%

%%%%%%%%%%%%%%%%%%%%%%%%%%%%%%%%%%%%%%%%%%%%%%%%%

\end{document}